\documentclass[twocolumn,superscriptaddress,amsmath,aps]{revtex4}

\usepackage{amssymb,mathrsfs,bm,color,times}
\usepackage{graphicx,color}
\usepackage{CJK}
\usepackage{bm}
\usepackage[hypertex]{hyperref}
\usepackage{amsmath}

\newcommand{\be}{\begin{equation}}
\newcommand{\ee}{\end{equation}}
\newcommand{\bea}{\begin{eqnarray}}
\newcommand{\eea}{\end{eqnarray}}
\newcommand{\bsube}{\begin{subequations}}
\newcommand{\esube}{\end{subequations}}

\newcommand{\Eq}[1]{Eq.\,(\ref{#1})}
\newcommand{\Eqs}[1]{Eqs.\,(\ref{#1})}

\newcommand{\la}{\langle}
\newcommand{\ra}{\rangle}

\newcommand{\beq}{\begin{equation}}
\newcommand{\eeq}{\end{equation}}
\newcommand{\beqn}{\begin{eqnarray}}
\newcommand{\eeqn}{\end{eqnarray}}

\newcommand{\nl}{\nonumber \\}

\newcommand{\bsub}{\begin{subequations}}
\newcommand{\esub}{\end{subequations}}

\begin{document}
\begin{CJK*}{GBK}{Song}

\title{Transport and fusion of Majorana zero modes
in the presence of nonadiabatic transitions}

\author{Qiongyao Wang}\thanks{These authors equally contributed to the work.}
\affiliation{Center for Joint Quantum Studies and Department of Physics,
School of Science, \\ Tianjin University, Tianjin 300072, China}
\author{Jing Bai}\thanks{These authors equally contributed to the work.}
\affiliation{Center for Joint Quantum Studies and Department of Physics,
School of Science, \\ Tianjin University, Tianjin 300072, China}

\author{Luting Xu}
\affiliation{Center for Joint Quantum Studies and Department of Physics,
School of Science, \\ Tianjin University, Tianjin 300072, China}

\author{Wei Feng}
\thanks{Corresponding author: fwphy@tju.edu.cn}
\affiliation{Center for Joint Quantum Studies and Department of Physics,
School of Science, \\ Tianjin University, Tianjin 300072, China}

\author{Xin-Qi Li}
\thanks{Corresponding author: xinqi.li@imu.edu.cn}
\affiliation{Center for Quantum Physics and Technologies, School of Physical Science and Technology,
Inner Mongolia University, Hohhot 010021, China}
\affiliation{Center for Joint Quantum Studies and Department of Physics,
School of Science, \\ Tianjin University, Tianjin 300072, China}

\date{\today}

\begin{abstract}
{\flushleft We perform simulations}
for transport and nontrivial fusion of Majorana zero modes
in topological superconducting quantum wires.
We uncover interesting behaviors of nonadiabatic transition
associated with the transport
through mini-gate-controlled multiple-segments modulations.
Owing to breaking of the initial fermion parity induced by nonadiabatic transitions,
a deviation from the statistics of outcomes of nontrivial fusion
arises and is analyzed.
Moreover, we develop a measurement scheme to infer the amount of
fermion parity breaking
and nonadiabatic transition probability to excited states,
based on the characteristic spectrum
of measurement current by a quantum-point-contact detector,
by measurement of the charge occupation dynamics in a fusion-outcome-probing quantum dot.
\end{abstract}

\maketitle

{\flushleft The nonlocal property}
of the Majorana zero modes (MZMs) and non-Abelian statistics obeyed
by them provide the foundation of application to topological
quantum computation \cite{Kit01,Kit03,Sar08,Ter15,Sar15,Opp20a}.
The nature of the non-Abelian statistics of the MZMs indicates that
braiding the MZMs can result in quantum state evolution in the manifold
of highly degenerate ground states \cite{Fish11,Opp12,Roy19,Han20,Fuji21}.
The non-Abelian nature of MZMs also indicates that
fusing a pair of MZMs can yield outcome of
either a vacuum, or an unpaired fermion (resulting in an extra charge)
\cite{Ali16,BNK20,NC22,Li24,Leij22,Zut20,Sau23}.
In literature this is usually termed as {\it nontrivial} fusion.
Actually, these braiding and fusing behaviors are closely related
to each other.
On one aspect (the fundamental aspect), the fusion with two outcomes
can serve as a demonstration of non-Abelian statistics,
since it indicates the quantum dimension $d>1$
(a degeneracy of the ground state manifold),
which leads to, after braiding the MZMs, a unitary evolution matrix
acting on the degenerate ground state manifold,
rather than to a scalar phase factor on a single non-degenerate state.
On the other aspect, even computationally speaking,
braiding is not needed if one can fuse arbitrary
sets of two and four zero-modes \cite{BNK20}.

In practice, typically, both braiding and fusion require moving/transporting
the MZMs in real space.
For instance, for braiding operations, the early and representative
scheme is quantum-adiabatically moving the MZMs by tuning a series
of electric gates, to drive different regions of the Majorana quantum
wire into the topological or non-topological regime
\cite{Fish11,Opp12,Tew11a,Plug17}.
This scheme is motivated by the fact that the MZMs will form at
the boundaries between the topological and non-topological regions.
With the progress of gating control techniques,
the Majorana moving schemes have gained
renewed interests in the past years \cite{Ali18,Roy19,Kel19,MacD20,Byr21}.
For fusion, as recently proposed in Ref.\ \cite{NC22}
and further analyzed in Ref.\ \cite{Li24},
nontrivial fusion of MZMs can be demonstrated
by fusing a pair of MZMs from two topological superconducting (TSC) wires
(each wire accommodating two MZMs at the ends).
In order to demonstrate the most interesting case of nontrivial fusion,
it is required to prepare
the initial pair states of MZMs with definite fermion parities,
and nonadiabatically move the MZMs together to fuse,
as hown schematically in Fig.\ 1(a).

\begin{figure}
  \centering
  \includegraphics[scale=0.5]{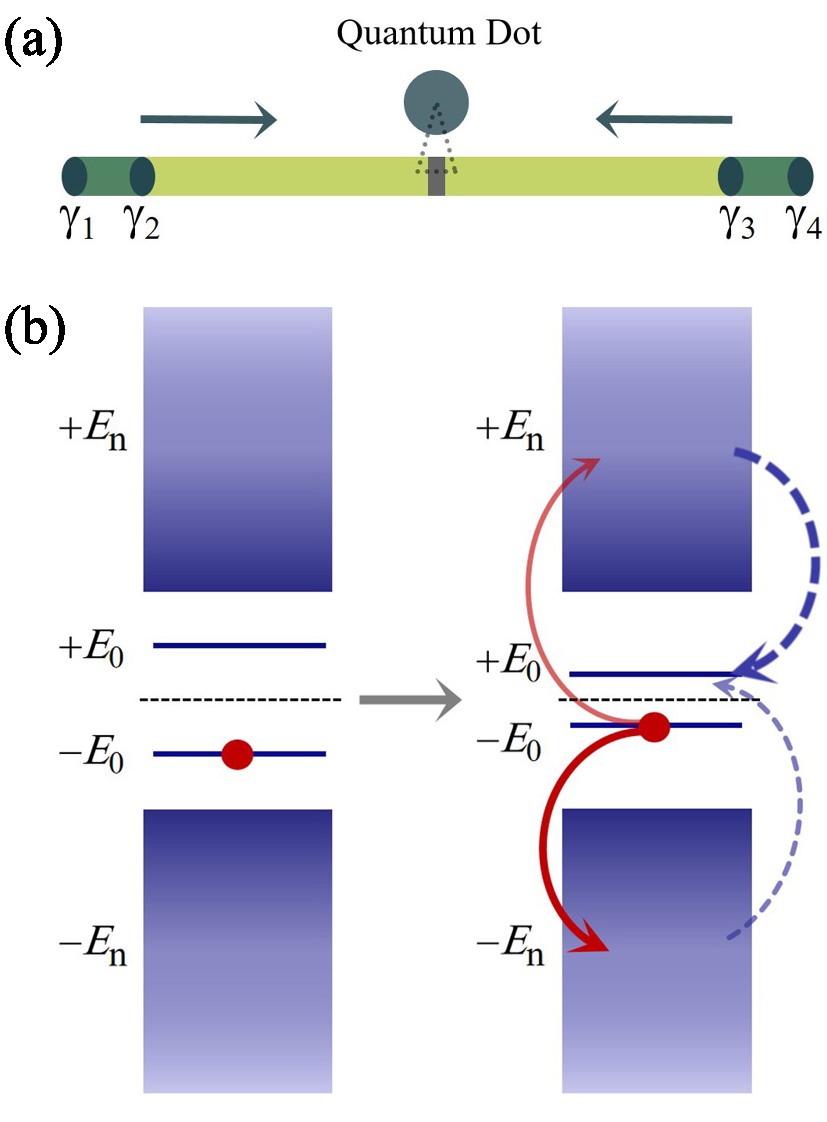}\\
\caption{
(a)
Schematic diagram of mini-gate controlled transport of MZMs
and nontrivial fusion of a pair of MZMs,
say, $\gamma_2$ and $\gamma_3$, from different Majorana pairs
with definite fermion parities (e.g. even parity) prepared in advance.
Based on the fusion rule $\gamma_2\times \gamma_3 = I + \psi$,
the fused MZMs would yield probabilistic outcomes
of vacuum $I$ and a regular fermion $\psi$.
A quantum dot is introduced to couple to the fused MZMs
for probing the fusion outcomes,
while the charge fluctuations in the quantum dot
can be detected by a nearby quantum-point-contact device (not shown
in this plot).
(b) BdG energy diagram of a TSC quantum wire (taking one of the
two wires shown here as an example, and omitting the labels
``L" and ``R" for the left and right wires).
The transport of the MZM is assumed to start with the
even parity state $|0\ra$,
which is described by occupation of the negative energy state
$|\psi_{-E_0}\ra$, as shown in the left panel.
In the right panel, a possible nonadiabatic transition
during moving the MZM is conceptually illustrated.    }
\end{figure}

For both braiding and fusion, in order to keep up the topological
protection, the quantum transport of MZMs should be adiabatically slow.
However, this may contradict other requirements, such as avoiding
the quasiparticle-poisoning decoherence.
Thus the effect of nonadiabatic transition constitutes an important
subject in the Majorana community
\cite{Ali18,Kel19,Sar11,Opp13,Shn13,Karz14,Fran17,Sau19,Tru23}.
Moreover, how to experimentally measure out the nonadiabatic transition
is another related important and interesting problem.
This has been addressed in an initial study in
Ref. \cite{Li22}, where a transport probe scheme (in terms of tunneling
spectroscopy of Majorana conductances) was analyzed.

In this work, following Ref.\ \cite{NC22},
and going drastically beyond Ref.\ \cite{Li24},
we perform simulations based on the lattice model of Rashba TSC quantum wires,
for the transport and fusion of MZMs,
starting with initial state of definite fermion parity (e.g., even parity).
Notice that, in Ref.\ \cite{Li24}, the effective low-energy MZMs
were directly used to analyze the fusion and probing dynamics,
thus the nonadiabatic effects caused by the moving
and their consequence
to the result of fusion and probing dynamics
cannot be accounted for. 
Specifically, in the present work, considering the
progress of mini-gate control technique, we will simulate the
gradual transport of MZMs by modulation of multiple
segments from non-topological to topological transitions, and
uncover interesting behaviors of nonadiabatic transition
associated with this type of modulation.
Being of great interest, owing to partly (weakly) switching to
an opposite-parity zero-energy-state, which is mediated by the
nonadiabatic transition to high energy excited-states,
we notice that the initial fermion parity will be broken by some amount.
Accordingly, the result will suffer some deviation
from the statistics of outcomes of nontrivial fusion.
Following Refs.\  \cite{NC22,Li24}, by considering the use of
a quantum-point-contact (QPC) detector and a quantum dot (QD)
to probe the outcomes of nontrivial fusion, very importantly,
we will develop a scheme to infer, {\it simultaneously}, both
the degree of fermion parity breaking
and nonadiabatic transition probability to excited states,
based on the characteristic spectrum of QPC measurement current.
The proposed scheme
of probing either the intrinsic fermion-parity breaking of MZMs,
or their nonadiabatic transition probability to excited states,
should be of great interest to the Majorana community,
considering that this type of studies are lacking in
literature, to the best of our knowledge.
\\

{\flushleft\it
Model and the Bogoliubov de-Gennes formulation}.---
For the various realizations
based on proximitized semiconductor/superconductor (SM/SC) hybrid structure,
the TSC quantum wire can be described
by a discretized lattice model as follows \cite{Ali18}
\bea\label{H-QW}
&& H_{\rm QW} = -\frac{W}{2} \sum_{i \sigma}(c_{i, \sigma}^{\dagger} c_{i+1, \sigma}
+c_{i+1, \sigma}^{\dagger} c_{i, \sigma})    \nl
&& +(W-\mu) \sum_{i \sigma} c_{i, \sigma}^{\dagger} c_{i, \sigma}
+V_{\rm z} \sum_{i \sigma \sigma^{\prime}} c_{i, \sigma}^{\dagger}
(\sigma^{z})_{\sigma \sigma^{\prime}} c_{i, \sigma^{\prime}}    \nl
&& +\frac{\alpha_{\rm so}}{2} \sum_{i \sigma \sigma^{\prime}}
\left[ c_{i, \sigma}^{\dagger}(i \sigma^{y})_{\sigma \sigma^{\prime}}
c_{i+1, \sigma^{\prime}}+c_{i+1, \sigma^{\prime}}^{\dagger}
(i \sigma^{y})_{\sigma \sigma^{\prime}} c_{i, \sigma} \right]    \nl
&& +\Delta \sum_{i \sigma}(c_{i, \uparrow} c_{i, \downarrow}
+c_{i,\downarrow}^{\dagger} c_{i, \uparrow}^{\dagger})  \,.
\eea
Following Ref.\ \cite{Ali18}, in the simulations of this whole work,
we choose the wire parameters (in a reduced arbitrary system of units) as:
the hopping energy $W=1$, the superconducing gap $\Delta=0.3$,
the Zeeman energy $V_{\rm z}=0.4$,
the spin-orbit-interaction (SOI) strength $\alpha_{\rm so}=0.3$,
and the chemical potentials
$\mu=\mu_{\rm T}=0$ and $\mu=\mu_{\rm nT}=-0.45$
for the topological and non-topological regimes.

For the purpose of numerical simulation of moving the MZMs
within the Bogoliubov de-Gennes (BdG) formalism,
the BdG Hamiltonian matrix $H_{\rm BdG}$
can be obtained through the following identity
\bea
H_{\rm QW} = \frac{1}{2} \hat{\Psi}^{\dagger} H_{\rm BdG} \hat{\Psi} \,,
\eea
where
$\hat{\Psi}=(c_{1\uparrow}\cdots c_{N\uparrow},c_{1\downarrow}
\cdots c_{N\downarrow},c_{1\uparrow}^{\dagger}\cdots c_{N\uparrow}^{\dagger},
c_{1\downarrow}^{\dagger}\cdots c_{N\downarrow}^{\dagger})^{T}$
is the so-called Nambu spinor.
The BdG Hamiltonian matrix $H_{\rm BdG}$ can be also understood as being constructed
under the basis of electron and hole states of the lattice sites,
i.e., $\{|e_{j\sigma}\ra,|h_{j\sigma}\ra; j=1,2,\cdots, N\}$.
Accordingly, the wavefunction of the TSC wire can be expressed as
$|\Psi(t)\ra=\sum_{j,\sigma} (u_{j\sigma}(t)|e_{j\sigma}\ra
+v_{j\sigma}(t)|h_{j\sigma}\ra)$,
which is the solution of the time-dependent BdG (TDBdG) equation
$i \partial_t  |\Psi\ra = H_{\rm BdG} |\Psi\ra$.
To identify the low-energy states (from which the MZMs are defined)
and nonadiabatic transitions to the excited Bogoliubov quasiparticle states,
it is useful to recast the wire state as
\bea\label{wire-WF}
&& |\Psi(t)\ra = \alpha_0 |\psi_{-E_0}\ra + \beta_0 |\psi_{+E_0}\ra  \nl
&& ~~~
+\sum_{n\neq 0} (\alpha_n |\psi_{-E_n}\ra + \beta_n |\psi_{+E_n}\ra ),
\eea
in which the instantaneous eigenstates are obtained through
$H_{\rm BdG}(t)|\psi_{\pm E_n}(t)\ra = \pm E_n(t)|\psi_{\pm E_n}(t)\ra$.
In the BdG formalism, the negative energy state $|\psi_{-E_n}\ra$
is the charge-conjugated counterpart
of the positive energy state $|\psi_{+E_n}\ra$,
holding the `particle' and `anti-particle' corresponding relation.

As schematically shown in Fig.\ 1(a),
we will consider the transport and fusion of the Majorana
modes $\gamma_2$ and $\gamma_3$, with an aim at demonstrating
the nontrivial fusion rule $\gamma_2\times \gamma_3 = I + \psi$.
This rule implies that the fused MZMs would yield probabilistic outcome
of the vacuum $I$ or a regular fermion $\psi$.
For this purpose, $\gamma_2$ and $\gamma_3$
should come from different Majorana pairs with definite fermion parities.
To be specific, the above nontrivial fusion rule can be examined
through the following state transformation \cite{Ali16,NC22}
\bea\label{FR}
|0_{12}0_{34}\ra
= \frac{1}{\sqrt{2}} (|0_{23}0_{14}\ra + i |1_{23}1_{14}\ra)  \,.
\eea
Here the regular fermion number states $|n_{12}n_{34}\ra$
and $|n_{23}n_{14}\ra$ are defined through different
combinations of the four Majorana modes.
For instance, $n_{12}$ is the particle number of the
regular fermion $f_{12}$, associated with the Majorana
modes $\gamma_1$ and $\gamma_2$.
Other particle number states are defined in the same way.
Therefore, in practice, we need to prepare the initial
state $|0_{12}0_{34}\ra$, and adiabatically move the Majorana
modes $\gamma_2$ and $\gamma_3$ together to fuse.
According to the proposal in Ref.\ \cite{NC22}, this can be realized as follows.
By means of mini-gate-voltage control,
first, move $\gamma_2$ and $\gamma_3$ to the ends of the two wires,
close to $\gamma_1$ and $\gamma_4$, respectively;
then, empty the possible occupations of the regular fermions $f_{12}$ and $f_{34}$
by introducing tunnel-coupled side quantum dots and modulating the dot energies
(while the quantum dots are also tunnel-coupled to outside reservoirs).
Starting with $|0_{12}0_{34}\ra$, move $\gamma_2$ and $\gamma_3$
from the two terminal sides back to the central part to fuse
such that $\epsilon_M\neq 0$.
In the basis of $n_{12}$ and $n_{34}$ occupations,
the fused state is still $|0_{12}0_{34}\ra$.
But in the basis of $n_{23}$ and $n_{14}$ occupations,
two possible outcomes with $n_{23}=0$ or 1 will be generated,
as shown above in \Eq{FR}.

Following Refs.\ \cite{NC22,Fuji21}, starting with $|0_{12}0_{34}\ra$,
but in order to simulate the possible nonadiabatic transitions,
we need to convert the description from the occupation number states
to the BdG positive and negative energy states.
We thus consider to simulate moving the MZMs $\gamma_2$
and $\gamma_3$ with the initial state $|\psi_{-E^{\rm L}_0},\psi_{-E^{\rm R}_0}\ra$.
This is actually the {\it dual counterpart} of starting with
$|1_{12}1_{34}\ra$, which should be converted into the initial
state $|\psi_{+E^{\rm L}_0},\psi_{+E^{\rm R}_0}\ra$, for the sake of moving
simulation within the BdG framework.
Here, we introduced the  superscripts ``${\rm L}$" and ``${\rm R}$"
to denote the left and right TSC wires.  \\
\\

{\flushleft\it Majorana transport and nonadiabatic transition}.---
The transport of MZMs can be realized
via the control of mini-gates as proposed in Ref.\ \cite{NC22}, i.e.,
sequentially changing the chemical potentials of segments of the quantum wire,
to realize transitions from the non-topological to the
topological regime.
Specifically, we assume the modulation of the chemical potential $\mu_j(t)$
of the $j_{\rm th}$ segment occurs according to
$\mu_j(t)=\{1-f[(t-t_j)/\tau]\}\mu_{\rm nT} + f[(t-t_j)/\tau]\mu_{\rm T}$,
with $f(s)$ a monotonically increasing function
and satisfying $f(0)=0$ and $f(1)=1$.
Following Ref.\ \cite{Ali18}, we assume $f(s)=\sin^2(s\pi/2)$.
For a $m$-segment modulation scheme with total time $T$,
the time of moving through each segment
is $\tau=t_{j+1}-t_j=T/m\equiv T_m$.
In this work, we will consider the following interesting problem:
for the same $T$,
what is the difference of nonadiabatic transition
between different choices of $m$ and $T_m$?
This type of information should be useful for experiments along this line.

\begin{figure}
  \centering
  \includegraphics[scale=0.5]{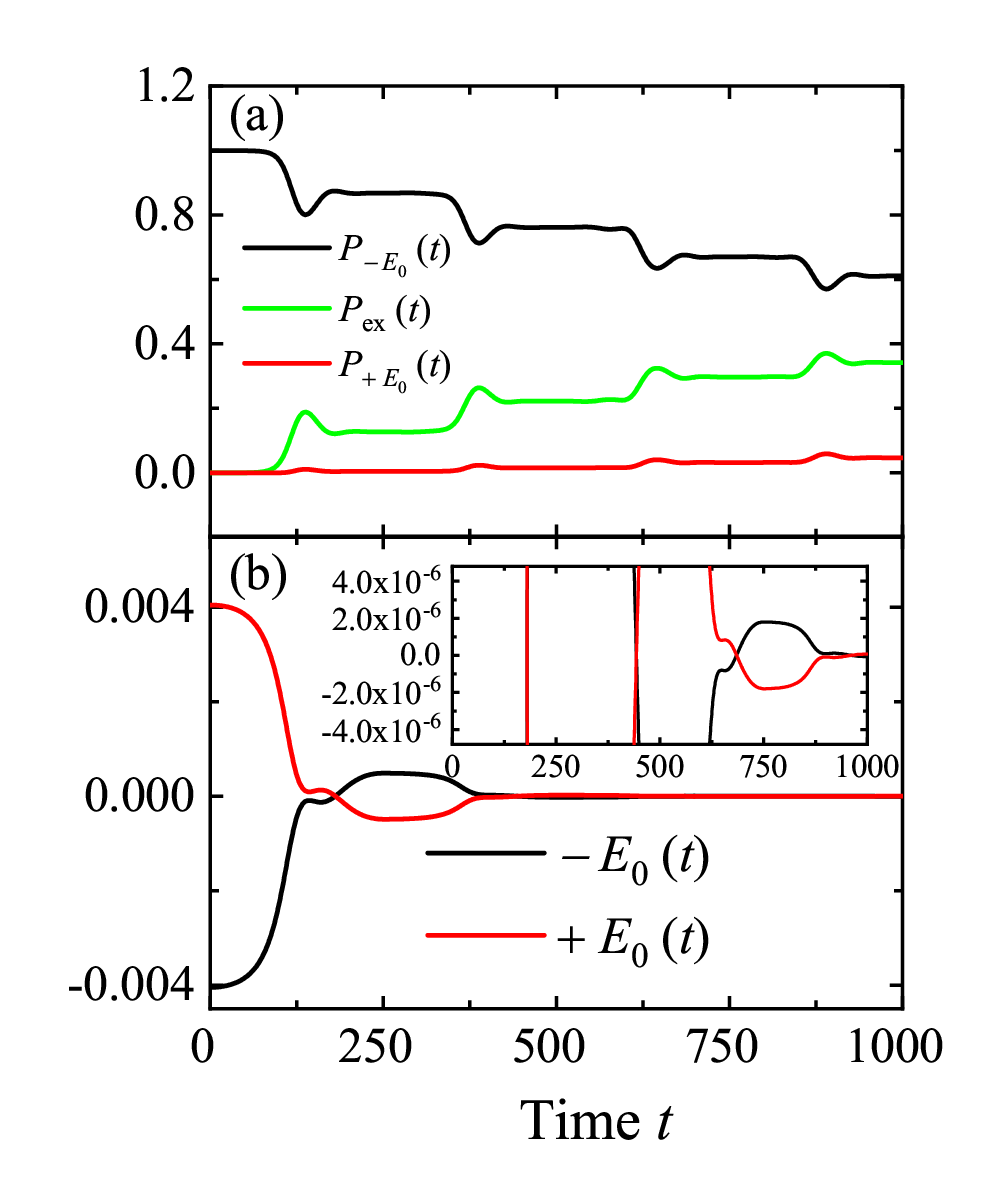}\\
  \caption{
Taking the 4-segments mini-gate controlled moving sequence
as an example, showing in (a) the dynamical evolution of
the occupation probabilities of the negative eigenstate
$|\psi_{-E_0}(t)\ra$, the positive eigenstate
$|\psi_{+E_0}(t)\ra$, and all the excited states;
and in (b) the energy diagram of the instantaneous eigenstates
$|\psi_{-E_0}(t)\ra$ and $|\psi_{+E_0}(t)\ra$.
The inset in (b) is the same result as in the main figure,
only using a smaller range of vertical coordinate, to enlarge the visual effect.
In order to determine $P_{-E_0}(t)$ and $P_{+E_0}(t)$,
it is necessary to properly track the evolution of
$|\psi_{-E_0}(t)\ra$ and $|\psi_{+E_0}(t)\ra$,
in the presence of Majorana energy oscillations
(positive and negative energy crossings).
Parameters of the TSC quantum wire used in the simulation
are referred to the main text below \Eq{H-QW}.     }
\end{figure}

In Fig.\ 2, we first display the basic behavior of a
nonadiabatic transition associated with the multiple-segments
modulation for moving the MZM in a single wire.
To be specific, we consider a quantum wire with $N=50$ lattice
sites and the modulation segment number $m=4$, with thus each
segment having 10 lattice sites.
In this work, we always consider moving the MZM in the wire
from a side topological segment with 10 lattice sites,
as schematically shown in Fig.\ 1(a).
In Fig.\ 2(a), we illustrate the dynamical evolution
of $P_{-E_0}(t)$, $P_{+E_0}(t)$, and $P_{\rm ex}(t)$
which are, respectively,
the occupation probabilities of the initial state $|\psi_{-E_0}\ra$,
the positive-lowest-energy state $|\psi_{+E_0}\ra$, and all excited states.
Here, $P_{\rm ex}(t)$ is simply defined through
$P_{\rm ex}(t)=\sum_{n\neq 0}[|\alpha_n(t)|^2+|\beta_n(t)|^2]$
from \Eq{wire-WF}, or equivalently,
through $P_{\rm ex}(t)= 1- [P_{-E_0}(t)+P_{+E_0}(t)]$.
Without loss of generality, here we consider moving a MZM in a single wire,
thus omitting the superscripts ``L" and ``R"
(which have been introduced to denote the left and right wires).
We find that when moving the MZM through each segment, via gate-voltage control,
nonadiabatic transition takes place,
which reduces $P_{-E_0}(t)$ and increases $P_{\rm ex}(t)$ and $P_{+E_0}(t)$.
In this context, we may first remark that
the nonadiabatic transition to the same type of BdG states,
i.e., the type of positive energy states
and the type of negative energy states,
is for the same reason of nonadiabatic transition
governed by quantum mechanics.
However, the nonadiabatic transition to the different type of
states involves the splitting or formation of a Cooper pair,
and is possible owing to the existence of a large number
of unpaired normal electrons in the superconductor.

Then, let us discuss the results of Fig.\ 2(a) in more detail.
The ``dip" and ``peak" behaviors of $P_{-E_0}(t)$ and $P_{\rm ex}(t)$
are owing to the closing of the superconducting gap,
during crossing from the non-topological to the
topological regime, thus suffering a relatively stronger nonadiabatic transition.
We also notice that the probability $P_{+E_0}(t)$ is much smaller
than $P_{\rm ex}(t)$.
This seemingly contradicts the {\it adiabatic condition} in
quantum mechanics,
since the energy difference of $|\psi_{+E_0}\ra$ from $|\psi_{-E_0}\ra$
is much smaller than the excited quasi-particle states from $|\psi_{-E_0}\ra$.
We may understand this important point as follows.
In Ref.\ \cite{Fuji21}, it was pointed out that
the particle-hole symmetry would prohibit {\it direct transition}
between the particle-hole symmetric eigenstates, i.e.,
$\Pi_{+E_n,-E_n}=i \la \psi_{+E_n}({\mathbf R})|\partial_t
[ \psi_{-E_n}({\mathbf R})] \ra = 0$,
while the Berry connection matrix is defined in more general as
$\Pi_{\pm E_n,\pm E_m}= i \la \psi_{\pm E_n}({\mathbf R})|
\partial_t [\psi_{\pm E_m}({\mathbf R})] \ra$,
with ${\mathbf R}$ denoting the time-dependent system parameters.
Then, we know that there is no direct nonadiabatic-transition-coupling
between $|\psi_{+E_0}\ra$ and $|\psi_{-E_0}\ra$.
The appearance of non-zero small $P_{+E_0}(t)$
is thus caused by an {\it indirect} nonadiabatic transition process,
i.e., from $|\psi_{-E_0}\ra$ to $|\psi_{+E_0}\ra$
mediated by the excited quasiparticle states.

We may remark that in order to determine $P_{-E_0}(t)$ and $P_{+E_0}(t)$,
it is necessary to track the time-dependent evolution of
the states $|\psi_{-E_0}\ra$ and $|\psi_{+E_0}\ra$.
In Fig.\ 2(b) we display the eigen-energies of these two states.
We find that the positive and negative energies interchange
after passing through a crossing point when the MZM moves through each segment.
This is the so-called {\it Majorana oscillation} phenomenon
\cite{Agu12,Sarm12,Loss13},
owing to the fact that, in realistic finite-length quantum wires,
the Majorana modes hybridize and the hybridization energy oscillates
as a function of the Zeeman energy, chemical potential, or wire length.
Properly addressing this issue is essential to obtain the results in Fig.\ 2(a),
otherwise sudden jumps of occupation would be wrongly obtained.

In Fig.\ 3 we display the nonadiabatic transition probability to excited states,
$P_{\rm ex}(T)$, for its $T$ (total moving time) dependence.
This qualitatively corresponds to the Landau-Zener tunneling behavior,
with an increase of $T$, say, causing the nonadiabatic transition
to become weaker.
Here we compare the results of three different transport schemes,
which are specified by $m=1$, 2, and 4 for, respectively,
the one-segment, two-segments, and four-segments transport schemes.
Interestingly, we find that the nonadiabatic transition is weaker
for the transport by means of multiple-segments modulation.

In Fig.\ 4, taking a few different values of $T$ as examples,
we detail what happens in the moving process for each moving scheme.
We see that, for the one-step moving, strong nonadiabatic transion
occurs around the superconducting-gap-closing place,
while for the multiple-segments moving scheme,
multiple nonadiabatic transions take place
near the gap-closing points in the moving through each segment,
as observed in the $m=2$ and 4 curves.
However, the summed probability of the multiple nonadiabatic transitions
becomes smaller with the increase of the segment numbers of mini-gate control.
This may partly explain the reason for the results in Fig.\ 3.

We may mention that,
our above result is only from
limited numerical simulations (for small number of segments).
The above result is basically consistent with that
revealed in Ref.\ \cite{Tru23},
where the transport of MZMs was simulated
using the Kitaev chain model
and analyzed by a generalized Landau-Zener transition formula.
In Ref.\ \cite{Tru23},
it was concluded that, in order to reduce nonadiabatic transition,
there exists an optimized choice of $m$
for the multiple segment control of MZM transport.
However, as shown in Ref.\ \cite{Ali18},
the Rashba TSC quantum wire is more complicated,
and the nonadiabatic transition in the Rashba TSC quantum wire
does not agree well with the Landau-Zener transition formula.
Whether or not the conclusion of Ref.\ \cite{Tru23}
is valid for the Rashba TSC quantum wire
seems an open question, which needs further systematic studies.

We may also mention that,
our results of the manipulation time $T$ dependence
display a large range of nonadiabatic transition behaviors.
However, the range of $T$ we simulated corresponds to moving velocities
below the critical velocity $v_c$ determined from the inequality of adiabatic condition,
as given by Eq.\ (38) and (43) in Ref.\ \cite{Shn13}.
Qualitatively speaking, after violating that condition,
nonadiabatic transition effect will be obvious.
However, such {\it inequality condition}
(which is used to define the critical velocity $v_c$)
does not mean something as the sharp condition (critical point)
in {\it phase transition} phenomena.
Below $v_c$, there exists also considerable nonadiabatic transition.
We also noticed that in Ref.\ \cite{Karz14},
it was concluded that when the domain-wall moving velocity
is larger than the {\it effective} light speed
(similar as the critical velocity $v_c$),
i.e., the Majorana mode is in a ``superluminal" motion,
the Majorana mode will become unstable,
say, with a transition taking place from localized to extended state.
However, this remarkable result
is likely needing further more careful and serious examinations,
especially by numerical simulations using various non-Dirac type electron models. 

\begin{figure}
  \centering
  \includegraphics[scale=0.46]{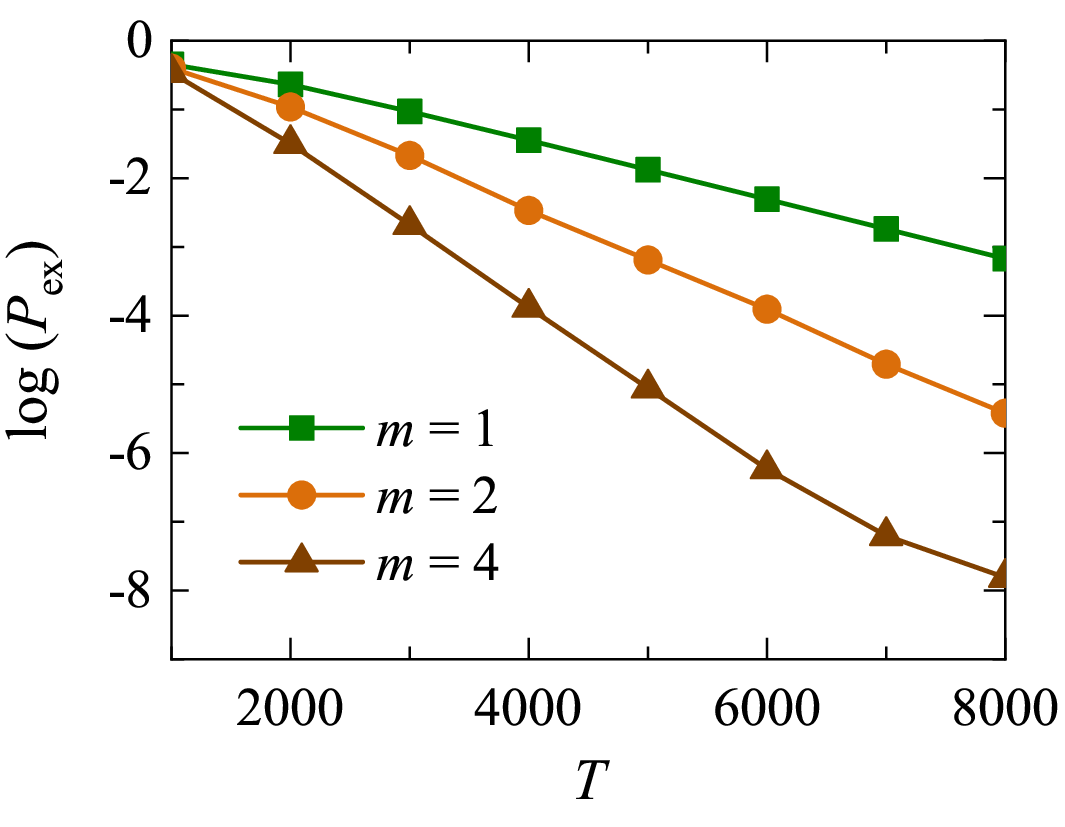}\\
  \caption{
Nonadiabatic transition probabilities to excited states,
$P_{\rm ex}=1-(P_{-E_0}+P_{+E_0})$,
{\it versus} the total moving time $T$
for different-$m$-segment moving schemes,
with $m$ defined as $T=mT_m$.
Parameters of the TSC quantum wire
are given in the main text below \Eq{H-QW}.      }
\end{figure}

\begin{figure}
  \centering
  \includegraphics[scale=0.6]{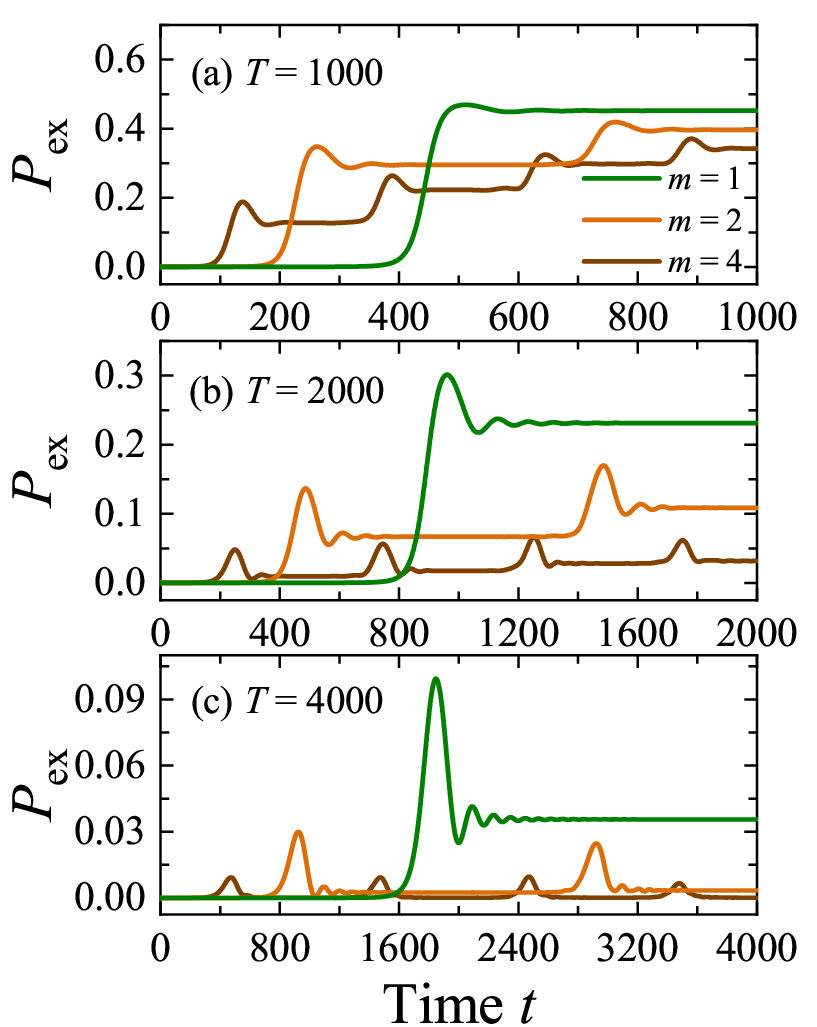}\\
  \caption{
Additional insight for the results shown in Fig.\ 3,
from the transient behaviors of the nonadiabatic transition probabilities
of different-$m$-segment moving schemes,
with $m$ defined as $T=mT_m$, while $T$ is the total moving time,
as exemplified by setting $T=1000$ in (a), 2000 in (b), and 4000 in (c).
Indeed, weaker nonadiabatic transition is found for moving
through more mini-gate controlled segments (larger $m$),
but with the same total time $T$.
Parameters of the TSC quantum wire
are given in the main text below \Eq{H-QW}.      }
\end{figure}

\begin{figure}
  \centering
  \includegraphics[scale=0.7]{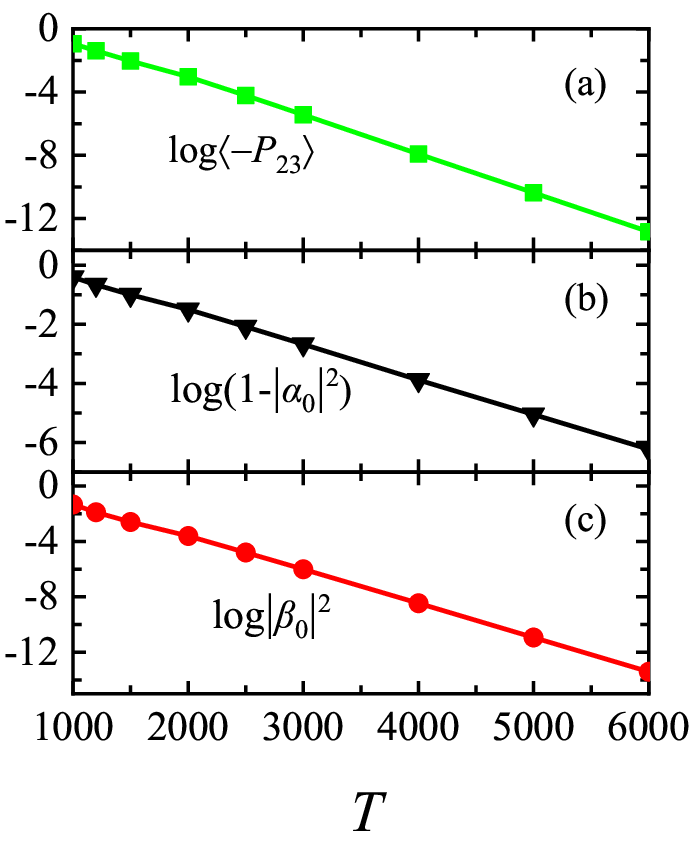}\\
  \caption{
Breaking of the initial fermion parity caused by nonadiabatic transition.
Frame (a) shows the degree of $\la P_{23}\ra\neq 0$ versus the moving time,
indicating a violation of the
statistics of nontrivial fusion outcomes,
with definite initial parity, which requires $\la P_{23}\ra = 0$ \cite{BNK20}.
Shown in (b) and (c) are the associated probabilities
of remaining in the initial state and going to the opposite parity state,
after moving the MZM in a single wire.
In essence, $\la P_{23}\ra\neq 0$ is a consequence of {\it quantum interference}
between the two different parity states (see the main text for details).
Parameters of the TSC quantum wire
are given in the main text below \Eq{H-QW}.    }
\end{figure}

{\flushleft\it Fermion parity breaking and its consequence}.---
After moving the MZMs $\gamma_2$ and $\gamma_3$ to the central part
(before fusing and coupling to the probing QD),
the nonadiabatic transition analyzed above
would render the states of the left and right wires as
$|\Psi_{\mu}(t)\ra = \alpha_{\mu} |\psi_{-E^{\mu}_0}\ra
+ \beta_{\mu} |\psi_{+E^{\mu}_0}\ra  + (\cdots)$.
Here $\mu$ denotes ``L" and ``R",
and $(\cdots)$ stands for the components of excited states.
Noting that Majorana fusion and probe-coupling to the QD
are dominantly taking place within the subspace of low-energy states,
the high-energy excited states will be gapped out
in the fusing and probing process.
Then the wire state after moving can be expressed as
\bea\label{WF-LR}
 |\Psi_{\rm LR}(0)\ra
&=& (\alpha_{\rm L}|0_{12}\ra + \beta_{\rm L}|1_{12}\ra)   \nl
&& \otimes(\alpha_{\rm R}|0_{34}\ra + \beta_{\rm R}|1_{34}\ra) \,.
\eea
Here the states of MZMs have been converted to
the number states of regular fermion occupation.
This product state has four components.
Indeed, the fusion rule of \Eq{FR}
accounts for the most interesting case
from the state $|0_{12} 0_{34}\ra$.

However, as pointed out in Ref.\ \cite{BNK20}, nontrivial fusion
can be reasonable for any initial state {\it with definite parity}.
Therefore, we carry out the following transformation rules \cite{Li24}
\bea\label{FR2}
|1_{12}1_{34}\ra
&=& \frac{1}{\sqrt{2}} (|0_{14}0_{23}\ra -i |1_{14}1_{23}\ra)  \,, \nl
|0_{12}1_{34}\ra
&=& \frac{1}{\sqrt{2}} (|1_{14}0_{23}\ra +i |0_{14}1_{23}\ra)   \,,  \nl
|1_{12}0_{34}\ra
&=& \frac{1}{\sqrt{2}} (|1_{14}0_{23}\ra -i |0_{14}1_{23}\ra)  \,.
\eea
Based on \Eqs{FR} and (\ref{FR2}), one can check that
the average of the parity $P_{23}=i\gamma_2\gamma_3$
is zero for any of the four states.
That is, for any (initial) state $|n_{12}n_{34}\ra$ with definite parity,
the fusion of $\gamma_2$ and $\gamma_3$ will result in
the statistical average
$\la P_{23}\ra = 0$, as formally proved in Ref.\ \cite{BNK20}.
Actually, this result reflects the key feature of nontrivial fusion
with equal weight outcomes of $I$ and $\psi$.

Accordingly, let us consider the statistical average of $P_{23}$
over the wire state $|\Psi_{\rm LR}(0)\ra$, given by \Eq{WF-LR}.
Let us reexpress this state in the basis $\{|n_{14} n_{23}\ra\}$.
Simple algebra yields
\bea\label{tot-0}
|\Psi_{\rm LR}(0)\ra
&=& \frac{1}{\sqrt{2}} (A_{+}|0_{14}\ra + B_{+}|1_{14}\ra )|0_{23}\ra  \nl
&&+ \frac{i}{\sqrt{2}} (A_{-}|1_{14}\ra + B_{-}|0_{14}\ra )|1_{23}\ra  \,.
\eea
Here we introduced $A_{\pm}=\alpha_{\rm L}\alpha_{\rm R} \pm \beta_{\rm L}\beta_{\rm R}$,
and $B_{\pm}=\alpha_{\rm L}\beta_{\rm R} \pm \beta_{\rm L}\alpha_{\rm R}$.
If we only consider the fusion of $\gamma_2$ and $\gamma_3$
(but not coupling to the QD),
the fusion just removes the energy degeneracy
between $|0_{23}\ra$ and $|1_{23}\ra$,
but renders the above state unchanged,
except a phase factor $e^{-i\epsilon_{\rm M} t}$
attached to $|1_{23}\ra$
(with $\epsilon_{\rm M}$ the coupling energy of $\gamma_2$ and $\gamma_3$).
This observation indicates that $\la P_{23}\ra$ is the same
before and after the fusion of $\gamma_2$ and $\gamma_3$:
\bea
&&\la P_{23}\ra  = -\frac{1}{2}(|A_+|^2+|B_+|^2)
        + \frac{1}{2}(|A_-|^2+|B_-|^2)  \nl
&& ~~~  = -2\, {\rm Re}[(\alpha_{\rm L}\alpha_{\rm R})^*(\beta_{\rm L}\beta_{\rm R})
        + (\alpha_{\rm L}\beta_{\rm R})^*(\beta_{\rm L}\alpha_{\rm R})] \,.
\eea
From this result, we see that nonzero $\la P_{23}\ra$ is caused
by the {\it nonadiabatic-transition-induced} mixing of $|1_{12}\ra$ and $|1_{34}\ra$
into the initially prepared left and wire states $|0_{12}\ra$ and $|0_{34}\ra$,
thus violating the definite parity condition of each wire.
In other words, the nonadiabatic transition would induce
a breaking of the initial fermion parity associated with the MZMs.
Furthermore, the result of nonzero $\la P_{23}\ra$
is caused by the interference, among the 4-MZMs states,
between the two even-parity states, and between the two odd-parity states.
As will become clear in the following,
this mechanism of leading to $\la P_{23}\ra\neq 0$ can be inferred
through a careful analysis for the QD occupation dynamics. \\
\\

{\flushleft\it Probing dynamics and information extraction}.---
Following Ref.\ \cite{NC22},
let us couple a nearby QD
(with a single energy level $\epsilon_{\rm D}$)
to the fused MZMs $\gamma_2$ and
$\gamma_3$, with coupling amplitudes $\lambda_2$ and $\lambda_3$.
In the picture of the regular fermion $f_{23}$,
the coupling Hamiltonian reads as
\bea
H' = (\lambda_{\rm N} d^{\dagger} f_{23}
+ \lambda_{\rm A} d^{\dagger} f_{23}^{\dagger})  + {\rm h.c.}  \,,
\eea
where $d^{\dagger}$ is the creation operator of the QD electron.
Physically, the first term describes the usual normal tunneling process
and the second term describes the Andreev process
owing to Cooper pair splitting and recombination.
The respective coupling amplitudes are associated with
$\lambda_2$ and $\lambda_3$ as
$\lambda_{\rm N,A}=\lambda_2 \pm i \lambda_3$.
We assume that the QD is initially prepared in an empty state,
$|0_{\rm d}\ra$, and the probe-coupling-caused state evolution is
described as
\bea\label{Uct}
U_{\rm c}(t) |0_{23}0_{\rm d}\ra
&=& \alpha_{\rm A}(t) |0_{23}0_{\rm d}\ra +  \beta_{\rm A}(t) |1_{23}1_{\rm d}  \ra  \,, \nl
U_{\rm c}(t) |1_{23}0_{\rm d}\ra
&=& \alpha_{\rm N}(t) |1_{23}0_{\rm d}\ra +  \beta_{\rm N}(t) |0_{23}1_{\rm d}  \ra  \,,
\eea
where $U_{\rm c}(t)$ is the coupling caused evolution operator.
Let us denote
$|\Psi_{\rm tot}(t)\ra=U_{\rm c}(t)(|\Psi_{\rm tot}(0)\ra \otimes |0_{\rm d}\ra)$.
The QD occupation is then obtained as
\bea\label{Pdt}
P_{\rm d}(t) &=& \la \Psi_{\rm tot}(t)|d^{\dagger} d
           |\Psi_{\rm tot}(t)\ra   \nl
&=& \frac{1}{2}(|\beta_{\rm A}(t)|^2 M_{\rm A} + |\beta_{\rm N}(t)|^2 M_{\rm N}) \,.
\eea
Here $M_{\rm A,N} = |A_{\pm}|^2+|B_{\pm}|^2$,
playing a role of {\it modification}
to the result of the ideal case (in the absence of
nonadiabatic transition),
i.e., $P_{\rm d}(t)=\frac{1}{2}(|\beta_{\rm A}|^2+|\beta_{\rm N}|^2)$.
Also, after a simple derivation, we obtain the
charge transfer probabilities into the QD through the two
channels (associated with the two fusion outcomes), as
\bea\label{Rabi}
|\beta_{\rm N,A}(t)|^2 = R_{\rm N,A} \sin^2(\Omega_{\rm N,A}  t)  \,.
\eea
Here we have defined
$R_{\rm N,A} \equiv |\lambda_{\rm N,A}|^2/\Omega^2_{\rm N,A}$,
while $\Omega_{\rm N,A} = \sqrt{\Delta^2_{\rm N,A}+|\lambda_{\rm N,A}|^2}$
(with $\Delta_{\rm N,A} = |\epsilon_{\rm D}\mp \epsilon_{\rm M}|/2$)
are the charge oscillation frequencies through the two channels.

Following Refs.\ \cite{NC22,Li24}, we consider monitoring
the charge occupation of the QD by a QPC detector.
In particular, we consider performing continuous weak measurement
to extract the dynamical information of charge occupation in the QD
from the output current power spectrum $S_I(\omega)$.
As shown in Refs.\ \cite{Kor01,Mil01,Li05,Opp20},
the structure of the current power spectrum is
$S_I(\omega)=S_0 + S_{\rm d}(\omega)$,
where $S_0$ is the frequency-free background noise
and $S_{\rm d}(\omega)$ is the information-contained part.
More specifically, $S_{\rm d}(\omega)$ is the Fourier transformation of
the correlation function $S_{\rm d}(\tau)$ of the dot occupation
($n_{\rm d}=d^{\dagger}d$)
\bea\label{Sd-tau}
S_{\rm d}(\tau)= {\rm Tr} [  n_{\rm d} e^{\mathcal{L}|\tau|} (n_{\rm d} \rho_{\rm st})] \,.
\eea
Here, as usual, we consider the current power spectrum of steady
state, $\rho_{\rm st}$.
The steady state
is defined by the stationary solution of the master equation
$\dot{\rho} = - i \left[H , \rho \right]
+ \kappa \mathcal{D}[n_{\rm d}]\rho \equiv \mathcal{L}\rho $,
where the Lindblad superoperator is defined as
$\mathcal{D}[x] \rho =x \rho_c x^{\dagger}
-\frac{1}{2}\{x^{\dagger}x,\rho_c \}$,
and $\kappa$ is the QPC measurement rate.

For each channel of charge oscillations
associated with $I$ or $\psi$ (i.e., $n_{23}=0$ or 1),
under the condition of weak-coupling measurement,
one can obtain the individual spectrum as
\cite{Li24,Kor01,Mil01,Li05,Opp20}
\bea\label{Lor}
&& S_{\rm d}^{(j)}(\omega)
\simeq \frac{\Delta_j^2}{4\Omega_j^2}
\frac{\kappa R_j/2}{\omega^2 + (\kappa R_j/2)^2}  \nl
&& ~~
 + \frac{|\lambda_j|^2}{8\Omega_j^2}
\frac{\frac{\kappa}{2} (1-\frac{R_j}{2})}
{(\omega-2\Omega_j)^2+[\frac{\kappa}{2} (1-\frac{R_j}{2})]^2 }  \,.
\eea
Here we use $j={\rm N, A}$
to denote the two charge transfer channels,
say, the normal tunneling and Andreev process.
From \Eq{Pdt}, we know that the two channels are independent of each other.
We thus expect the total spectrum to be a weighted sum of the individual
$S^{\rm (N)}_{\rm d}(\omega)$ and $S^{\rm (A)}_{\rm d}(\omega)$,
as follows
\bea\label{Sd-omega}
S_{\rm d}(\omega)=\frac{1}{2}
\left[M_{\rm N} S^{\rm (N)}_{\rm d}(\omega)
+ M_{\rm A} S^{\rm (A)}_{\rm d}(\omega)\right] \,.
\eea
Indeed, as shown in Fig.\ 6, a full agreement is demonstrated
between this analytic result [together with \Eq{Lor}]
and the numerical result from computation
using the full states, \Eqs{tot-0} and (\ref{Uct}),
and the master-equation-based \Eq{Sd-tau}.

\begin{figure}
\centering
\includegraphics[scale=0.45]{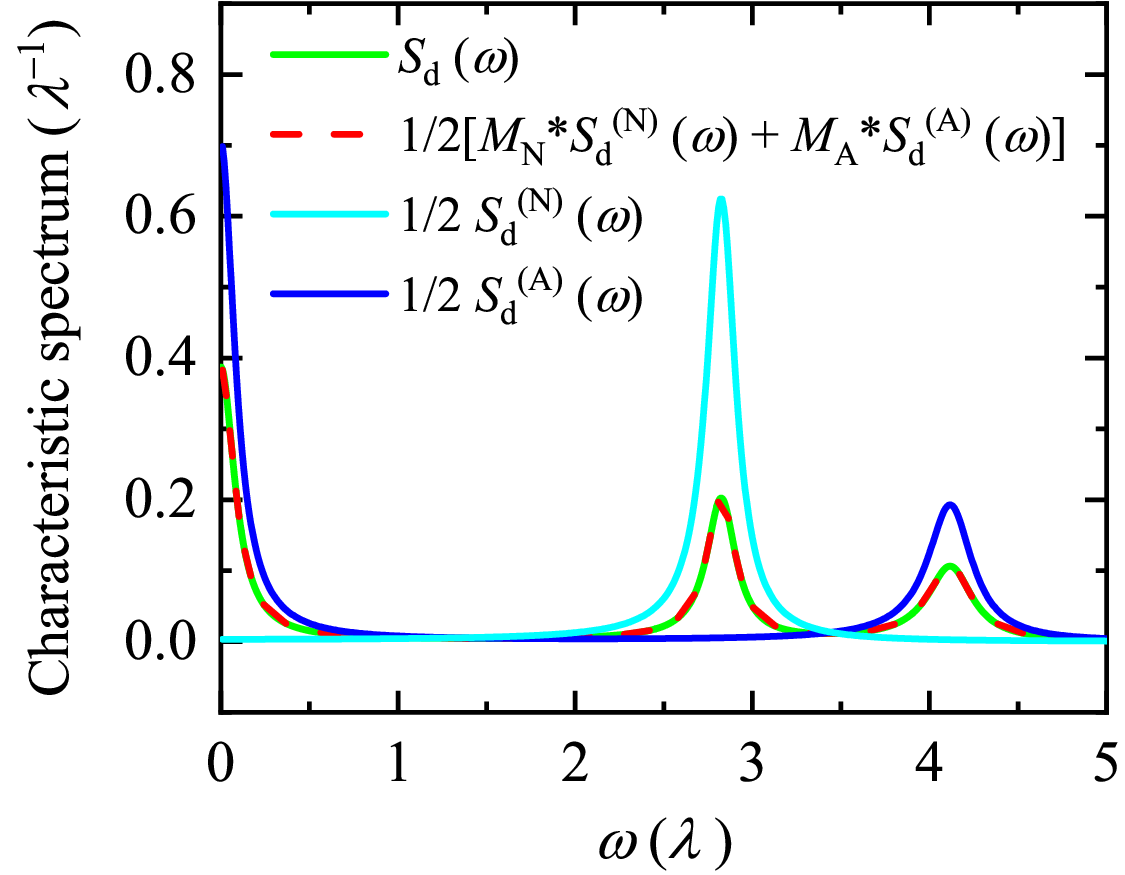}\\
\caption{The characteristic spectrum, $S_{\rm d}(\omega)$,
the Fourier transform of the QD-occupation correlation function,
singled out from the QPC output current spectrum of continuous
weak measurement.
The exact numerical result (solid green line) is compared with
the theoretically approximated result (dashed red line)
based on \Eqs{Lor} and (\ref{Sd-omega}), and perfect agreement is
demonstrated.
From the location of the two peaks, one obtains
the characteristic frequencies $\Omega_{\rm N}$ and $\Omega_{\rm A}$,
of charge oscillations associated with the fusion outcomes $\psi$ and $I$.
Importantly, as analyzed in detail in the main text,
from the heights of the two peaks,
one can infer the degree of breaking of the initial
fermion parity (definite parity) caused by nonadiabatic
transitions, which is characterized by a nonzero
$\la P_{23} \ra$, and infer the probability $P_{\rm ex}$
of nonadiabatic transition to all the excited states.
In addition to the parameters of the TSC quantum wire
given below \Eq{H-QW}, the numerical results shown in this
plot are from a choice of other reduced parameters as:
$\lambda_2=\lambda_3=\lambda$,
$\epsilon_{\rm D}=\epsilon_{\rm M}=1.5\lambda$, and $\kappa=0.4\lambda$.
Notice that, in order to guarantee only the subgap
low-energy-states involving the charge transfer dynamics,
the condition $\lambda  \ll  \Delta$ is required.}
\end{figure}

From \Eqs{Lor} and (\ref{Sd-omega}), as shown in Fig.\ 6,
we know that in $S_{\rm d}(\omega)$, two characteristic peaks
will appear at $\omega=2\Omega_{\rm N}$ and
$2\Omega_{\rm A}$. The heights of the two peaks
are given by $h_{\rm N}= M_{\rm N} h_{\rm N}^{(0)}/2$
and $h_{\rm A}= M_{\rm A} h_{\rm A}^{(0)}/2$,
while $h_{\rm N}^{(0)}$ and $h_{\rm A}^{(0)}$ are, based on \Eq{Lor}, known as
\bea\label{hj0}
h_j^{(0)} = \frac{1}{4\kappa} \frac{R_j}{1-R_j/2}  \,.
\eea
Then, one can extract the {\it modification factors}
from the heights of the peaks through
$M_{\rm N}=2h_{\rm N}/h^{(0)}_{\rm N}$ and $M_{\rm A}=2h_{\rm A}/h^{(0)}_{\rm A}$.
Very importantly, one can check the following relation
\bea
M_{\rm N}-M_{\rm A}= 2 \la P_{23}\ra \,.
\eea
This indicates that we can infer
the deviation from the statistics of outcomes of nontrivial fusion
in ideal case, which requires $\la P_{23}\ra=0$.
Also, in the symmetric case, say,
$|\alpha_{\rm L}|=|\alpha_{\rm R}|=|\alpha_0|$
and $|\beta_{\rm L}|=|\beta_{\rm R}|=|\beta_0|$,
we find another important relation
\bea
M_{\rm N}+M_{\rm A} = 2 (|\alpha_0|^2+|\beta_0|^2)^2 \,.
\eea
From it, one can infer the nonadiabatic transition probability
to excited states in each wire,
i.e., $P_{\rm ex}=1-(|\alpha_0|^2+|\beta_0|^2)$.

In practice, rather than measuring the various parameters to determine
$h_{\rm N}^{(0)}$ and $h_{\rm A}^{(0)}$ according to \Eq{hj0},
one can obtain them in a more practical manner as follows.
As initializing the empty occupation $n_{12}=0$ and $n_{34}=0$
of the regular fermions $f_{12}$ and $f_{34}$,
we can assume the same way to empty the occupation of the regular fermion $f_{23}$
associated with the fused MZMs $\gamma_2$ and $\gamma_3$,
by introducing an additional tunnel-coupled side quantum dot
(electron-mediating-QD)
and modulating the dot energy level $\widetilde{\epsilon}_{\rm D}$
in resonance with the Majorana energy $\epsilon_{\rm M}$,
while the electron-mediating-QD
is tunnel-coupled to an outside reservoir
with Fermi level lower than $\widetilde{\epsilon}_{\rm D}$.
This can ensure $n_{23}=0$. Starting with it,
perform the measurement of steady-state current power spectrum,
one can obtain the single peak height $h_{\rm A}^{(0)}$ at $\omega=2\Omega_{\rm A}$.
In order to obtain $h_{\rm N}^{(0)}$,
after ensuring $n_{23}=0$ as explained above,
one can input an electron from the reservoir into an occupation
of the regular fermion $f_{23}$ through the electron-mediating-QD,
by increasing the Fermi level of the reservoir above $\widetilde{\epsilon}_{\rm D}$.
Then, starting with $n_{23}=1$
and performing the steady-state spectrum measurement,
one can obtain the single peak height $h_{\rm N}^{(0)}$ at $\omega=2\Omega_{\rm A}$.
Finally, we remark that, in order to realize the definite occupation result
$n_{23}=0$ or $n_{23}=1$ as described above,
Andreev process should be suppressed.
This requires a weak coupling of the regular fermion $f_{23}$
to the electron-mediating-QD, i.e., with the coupling strength
much smaller than $\epsilon_{\rm M}$ and $\widetilde{\epsilon}_{\rm D}$.  \\
\\
{\flushleft\it Summary and discussion}.---
For the purpose of demonstrating the nontrivial fusion of MZMs,
which requires a preparation of
initial state of Majorana pairs with definite fermion parity
and moving the MZMs (to be fused) together from distant locations,
we performed simulations of moving a MZM in a TSC quantum wire.
In particular, we simulated the gradual moving of a MZM through modulations
of multiple segments from non-topological to topological transitions,
and displayed interesting behaviors of the nonadiabatic transition.
We also analyzed in detail the result of intrinsic fermion-parity breaking of MZMs,
by identifying its origin of being induced/mediated by the
nonadiabatic transition to high energy excited states,
and analyzed its consequence to the result of
nontrivial fusion of two MZMs.
Moreover,
and most importantly,
we developed a scheme to simultaneously infer both
the degree of fermion parity breaking
and nonadiabatic transition probability to excited states,
based on the characteristic spectrum of measurement current,
by using the locations and heights of two characteristic peaks.
We are not aware of similar studies in literature, and this
probing protocol should therefore be of great interest to the Majorana community.

About the deviation from the statistics of outcomes of nontrivial fusion,
owing to fermion parity breaking
during moving the MZM in each single TSC quantum wire,
we may add some more discussions as follows.
In our simulation of moving the Majorana mode,
we only consider two Majorana modes in a single wire
(e.g. $\gamma_1$ and $\gamma_2$).
Associated with the fermion parity breaking,
the state transfer sequence is:
$|0_{12}\ra \rightarrow  |1_n\ra \rightarrow |1_{12}\ra$.
Here, to be more clear,
we use the number-state representation.
In the first process, a Cooper pair is split into two electrons,
with one electron being excited to the gap-above quasiparticle state $|1_n\ra$,
while another electron being submerged into the sea of background normal electrons.
In the second process,
the electron is transitioned from the quasiparticle state
to the sub-gap (near zero-energy) state.
As a result, the occupation of
the sub-gap state of a single wire (e.g. the left wire) is
$\alpha_{\rm L}|0_{12}\ra + \beta_{\rm L}|1_{12}\ra$.
For the two wires,
the total state is a product of the two wire states, as shown by \Eq{WF-LR}.
Then, owing to the appearance of $|1_{12}\ra$ and $|1_{34}\ra$
(i.e. violation of the fermion parity of each single wire),
the quantum average of the parity operator $P_{23}=i\gamma_2\gamma_3$
becomes nonzero, $\la P_{23}\ra \neq 0$,
which differs from the result $\la P_{23}\ra = 0$,
i.e., the prediction for the statistics of the nontrivial fusion outcomes \cite{BNK20}.

We notice that in Ref.\ \cite{Kel19}, the 4-MZMs-qubit is considered
for possible qubit loss (escape from the subspace of qubit states),
bit-flit error, and phase-flip error,
mediated by the nonadiabatic transition (through the excitation spectrum),
owing to moving the Majorana mode and/or noise-caused fluctuations.
For the bit flip between the logic states
$|0_{12} 0_{34}\ra$ and $|1_{12} 1_{34}\ra$,
an effective coupling between $\gamma_2$ and $\gamma_3$
may be explored to describe it, using the coupling Hamiltonian
$H'=i\epsilon_{23}\gamma_2 \gamma_3$,
which is similar to the parity operator $P_{23}=i\gamma_2\gamma_3$.
Actually, in Ref.\ \cite{Kel19}, the qubit flip error
$|0_{12} 0_{34}\ra \rightarrow |1_{12} 1_{34}\ra$,
mediated by the nonadiabatic transition through the gap-above excited states,
experiences the following state transfer sequence:
$|0_{12} 0_{34}\ra |0_n\ra \rightarrow |1_{12} 0_{34}\ra|1_n\ra
\rightarrow |1_{12} 1_{34}\ra|0_n\ra$,
where $|1_n\ra$ and $|0_n\ra$ correspond to
the excited state $|\psi_{+E_n}\ra$ occupied and unoccupied.
The underlying physics of the first process
is associated with splitting of a Cooper pair,
while the second process is a usual quantum transition
from $|\psi_{+E_n}\ra$ to $|\psi_{+E_0}\ra$.

The essential difference between the result
of bit-flip error in Ref.\ \cite{Kel19} and ours is:
for the flip error of the 4-MZMs qubit,
$|0_{12} 0_{34}\ra \rightarrow |1_{12} 1_{34}\ra$,
the total fermion parity is conserved;
while in our case, the result $|0_{12}\ra \rightarrow  |1_{12}\ra$
indicates a violation of the fermion parity,
in a {\it closed} single wire.
In our opinion, the latter effect is of more interest.
In practice, this effect can be measured along the analysis in this work.
In order to restrict the nonadiabatic transition occurred in each single wire,
the two wires should be interrupted at the center
by a gate-controlled potential barrier [see Fig.\ 1(a)],
before the two MZMs (to be fused)
are moved close to each other (during the moving process).
The fusion can be realized
by lowering the barrier after the two MZMs are moved close to each other.    

\vspace{0.5cm}
{\flushleft\it Acknowledgements.}---
This work was supported by the
National Key Research and Development Program of China
(No.\ 2017YFA0303304) and the NNSF of China (Nos.\ 11675016, 11974011 \& 61905174).


\end{CJK*}
\end{document}